\documentstyle[emulateapj,psfig]{article}

\makeatletter

\newenvironment{inlinefigure}{%
\def\@captype{figure}%
\noindent\begin{minipage}{0.999\linewidth}\begin{center}}
{\end{center}\end{minipage}\smallskip}
\makeatother

\lefthead{Barger et al.}

\begin{document}
\title{The Number Density of Intermediate and High
Luminosity Active Galactic Nuclei at $z\sim 2-3$}
\author{
A.~J.~Barger,$\!$\altaffilmark{1,2,3}
L.~L.~Cowie$\!$\altaffilmark{3}
}

\altaffiltext{1}{Department of Astronomy, University of Wisconsin-Madison,
475 North Charter Street, Madison, WI 53706}
\altaffiltext{2}{Department of Physics and Astronomy,
University of Hawaii, 2505 Correa Road, Honolulu, HI 96822}
\altaffiltext{3}{Institute for Astronomy, University of Hawaii,
2680 Woodlawn Drive, Honolulu, HI 96822}

\slugcomment{In press at The Astrophysical Journal (10 December 2005 issue)}

\begin{abstract}
We use the combination of the 2~Ms {\em Chandra\/} X-ray image,
new $J$ and $H$ band images, and the {\em Spitzer\/} IRAC and MIPS 
images of the {\em Chandra\/} Deep Field-North to obtain high 
spectroscopic and photometric redshift completeness of high 
and intermediate X-ray luminosity sources in the redshift interval 
$z=2-3$. We measure the number densities of $z=2-3$ active 
galactic nuclei (AGNs) and broad-line AGNs in the rest-frame
$2-8$~keV luminosity intervals 
$10^{44}-10^{45}$ and $10^{43}-10^{44}$~ergs~s$^{-1}$ 
and compare with previous lower redshift results. 
We confirm a decline in the number densities 
of intermediate-luminosity sources at $z>1$. We also measure 
the number density of $z=2-3$ AGNs in the luminosity interval 
$10^{43}-10^{44.5}$ and compare with previous low and high-redshift 
results. Again, we find a decline in the number densities at $z>1$. 
In both cases, we can rule out the hypothesis that the number
densities remain flat to $z=2-3$ at above the $5\sigma$ level.
\end{abstract}

\keywords{cosmology: observations --- galaxies: active --- 
galaxies: evolution --- galaxies: formation --- galaxies: distances
and redshifts}

\section{Introduction}
\label{secintro}

Low-redshift hard X-ray luminosity functions have been well 
determined from the combination of highly spectroscopically 
complete deep and wide-area {\em Chandra\/} X-ray surveys. 
At $z<1.2$, the hard X-ray luminosity 
functions for active galactic nuclei (AGNs) of all spectral 
types and for broad-line AGNs alone are both
well described by pure luminosity evolution, with $L_\ast$
evolving as $(1+z)^{3.2}$ and $(1+z)^{3.0}$,
respectively (Barger et al.\ 2005). AGNs decline in 
luminosity by almost an order of magnitude over this redshift
range.

Barger et al.\ (2005) compared directly their broad-line AGN
hard X-ray luminosity functions with the optical QSO luminosity
functions from Croom et al.\ (2004) and found that the bright 
end luminosity functions agree extremely well at all redshifts. 
However, the optical QSO luminosity functions do not probe faint 
enough to see the downturn in the broad-line AGN hard X-ray 
luminosity functions at low luminosities and may be missing some 
sources at the very lowest luminosities to which they probe.

The Croom et al.\ (2004) pure luminosity evolution is slightly
steeper than that of Barger et al.\ (2005), but within the 
uncertainties, the two determinations are consistent over
the $z=0-1.2$ redshift interval. To investigate whether
pure luminosity evolution continues to hold at higher redshifts, 
Barger et al.\ (2005) used the Croom et al.\ (2004) 
evolution law (which was fitted over the wider redshift range 
$z=0.3-2.1$) to correct all of their broad-line AGN hard X-ray
luminosities to the 
values they would have at $z=1$. They then computed the 
broad-line AGN hard X-ray luminosity 
functions over the wide redshift intervals $z=0.2-0.7$, $0.7-1.5$, 
and $1.5-2.5$. Barger et al.\ (2005) found that the lower redshift 
luminosity functions matched each other throughout the luminosity 
range, while the highest redshift luminosity function matched 
the lower redshift functions only at the bright end, where the 
optical QSO determinations were made. They therefore concluded 
that there are fewer intermediate X-ray luminosity broad-line 
AGNs in the $z=1.5-2.5$ redshift interval, and hence that the 
pure luminosity evolution model cannot be carried reliably to 
the higher redshifts. This was consistent with other analyses 
made of less complete samples of all spectral types together, 
which had found evidence for peaks and subsequent declines in 
the number densities of both intermediate-luminosity
(e.g., Cowie et al.\ 2003; Barger et al.\ 2003a; Hasinger 2003; 
Fiore et al.\ 2003; Ueda et al.\ 2003) and high-luminosity 
(Silverman et al.\ 2005) sources, with the intermediate-luminosity 
sources peaking at lower redshifts than the high-luminosity 
sources.

Recently, Nandra et al.\ (2005; hereafter, N05) 
have questioned the evidence for a decline in the number densities 
of intermediate-luminosity AGNs at $z>1$. By combining deep X-ray data 
with the Lyman break galaxy (LBG) surveys of Steidel et al.\ (2003),
they argue that the number densities of intermediate-luminosity
AGNs are roughly constant with redshift above $z=1$.

Only with extremely deep X-ray data and highly complete redshift 
identifications can we address the above controversy about a 
decline. In this paper, we measure the $z=2-3$ number densities 
of both high and intermediate X-ray luminosity sources in the
2~Ms {\em Chandra\/} Deep Field-North (CDF-N). In addition to our 
highly complete spectroscopic redshift identifications, we also 
use the recently released {\em Spitzer\/} Great Observatories 
Origins Deep Survey-North IRAC and MIPS data,
in combination with deep $J$ and $H$ band data obtained from 
the ULBCAM instrument on the University of Hawaii's 2.2~m telescope,
to estimate accurate infrared (IR) photometric redshifts 
for the X-ray sources that could not be spectroscopically or 
optically photometrically identified. We assume $\Omega_M=0.3$, 
$\Omega_\Lambda=0.7$, and $H_0=70$~km~s$^{-1}$~Mpc$^{-1}$ throughout.

\section{Data}
\label{secobs}

The deepest X-ray image is the $\approx2$~Ms CDF-N exposure 
centered on the Hubble Deep Field-North taken with the ACIS-I
camera on {\em Chandra\/}. Alexander et al.\ (2003)
merged samples detected in seven X-ray bands to form a catalog of 
503 point sources over an area of 460~arcmin$^2$. Near the aim 
point, the limiting fluxes are 
$\approx 1.5\times 10^{-17}$~ergs~cm$^{-2}$~s$^{-1}$ ($0.5-2$~keV) and 
$\approx 1.4\times 10^{-16}$~ergs~cm$^{-2}$~s$^{-1}$ ($2-8$~keV). 

The ground-based, wide-field optical data of the CDF-N are 
summarized in Capak et al.\ (2004), and the {\em HST\/}
ACS GOODS-N data are detailed in Giavalisco et al.\ (2004).
The {\em Spitzer\/} IRAC and MIPS data are from the 
{\em Spitzer\/} Legacy data products
release and are presented in Dickinson et al.\ (2005).
The new ULBCAM $J$ and $H$ band data, which cover the whole CDF-N 
area to $1\sigma$ depths of $25.2$ in $J$ and $24.2$ in $H$
(these are measured in $3''$ apertures and corrected to total 
magnitudes), are described in Trouille et al.\ (2005). 
All magnitudes in this paper are in the AB system.
Most of the spectroscopic redshifts for the X-ray sources are 
from Barger et al.\ (2002, 2003b, 2005), but we also obtained 
some additional redshifts with the DEIMOS spectrograph on the 
Keck~II telescope during the Spring 2005 observing season.
One spectroscopic redshift ($z=2.578$) is taken from 
Chapman et al.\ (2005).

\subsection{X-ray Incompleteness}

We consider in our analysis two restricted, uniform, flux-limited 
X-ray samples. The most important consideration in choosing our 
soft X-ray flux limits was that there be no significant X-ray 
incompleteness. In Figure~\ref{fig1}, we show $0.5-2$~keV flux 
versus off-axis radius for the full 2~Ms CDF-N X-ray sample of
Alexander et al.\ (2003) ({\em squares and diamonds\/}).
We use small squares to denote sources that are detected in the
2~Ms survey but not in the 1~Ms survey (Brandt et al.\ 2001).
For the bright subsample, we consider sources in a $10'$ radius 
region with fluxes above 
$1.7\times 10^{-15}$~ergs~cm$^{-2}$~s$^{-1}$ ($0.5-2$~keV),
and for the deep subsample, we consider sources in an $8'$ 
radius region with fluxes above 
$1.7\times 10^{-16}$~ergs~cm$^{-2}$~s$^{-1}$ ($0.5-2$~keV). 
We show these flux and radius limits with solid lines.
The two subsamples comprise 160 sources, 8 of which are stars.

We can see from Figure~\ref{fig1} that the flux limits of our 
subsamples are well above the flux limit of the 2~Ms survey 
({\em dotted curve\/}) at all radii, suggesting that there 
should be very little X-ray incompleteness in our sample. 
Indeed, when we consider the 160 sources in the 2~Ms exposure 
that lie within our two subsamples, we find that only five sources 
were not already detected in the 1~Ms exposure. If we were using 
the 1~Ms catalog in our analysis rather than the 2~Ms catalog, 
then this would correspond to a 3\% incompleteness. With 
the 2~Ms exposure, we expect X-ray incompleteness to be negligible.

%
%
\begin{inlinefigure}
\psfig{figure=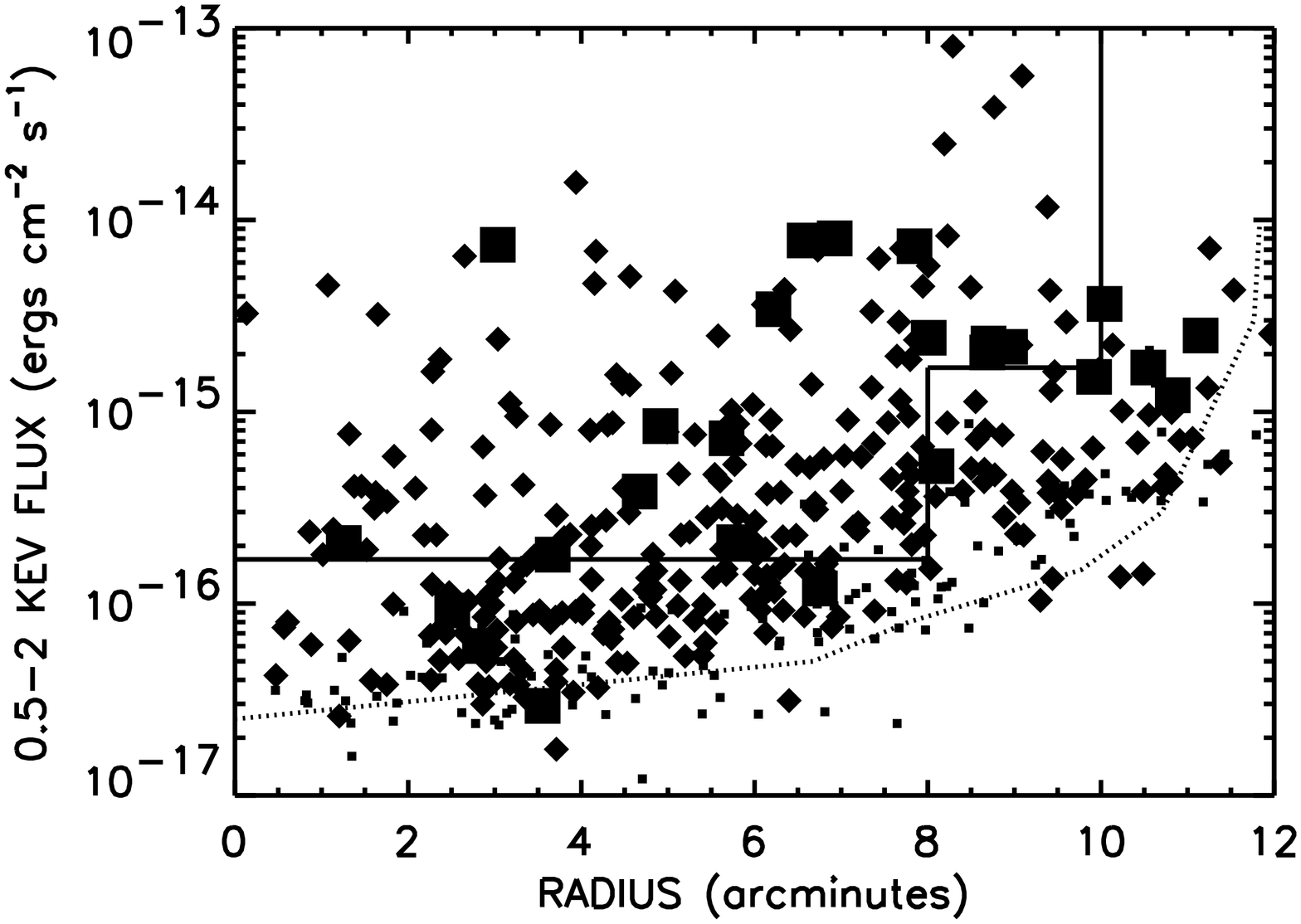,angle=90,width=3.5in}
\vspace{6pt}
\figurenum{1}
\caption{
$0.5-2$~keV flux vs. off-axis radius for the full 2~Ms CDF-N sample
of Alexander et al.\ (2003) ({\em squares and diamonds\/}).
The flux and radius limits of our bright and faint subsamples are
denoted by the solid lines, and the 2~Ms flux limit is denoted by
the dotted curve. Sources detected in the 2~Ms survey but not in
the 1~Ms survey (Brandt et al.\ 2001) are denoted by small squares.
Only five such sources lie within our subsamples.
Sources with spectroscopic redshifts in the interval $z=2-3$ are
denoted by large squares. Our flux and radius limits were chosen
without reference to the distribution of the $z=2-3$ sources.
\label{fig1}
}
\addtolength{\baselineskip}{10pt}
\end{inlinefigure}

\subsection{Redshift Interval}

\begin{figure*}
\centerline{\psfig{figure=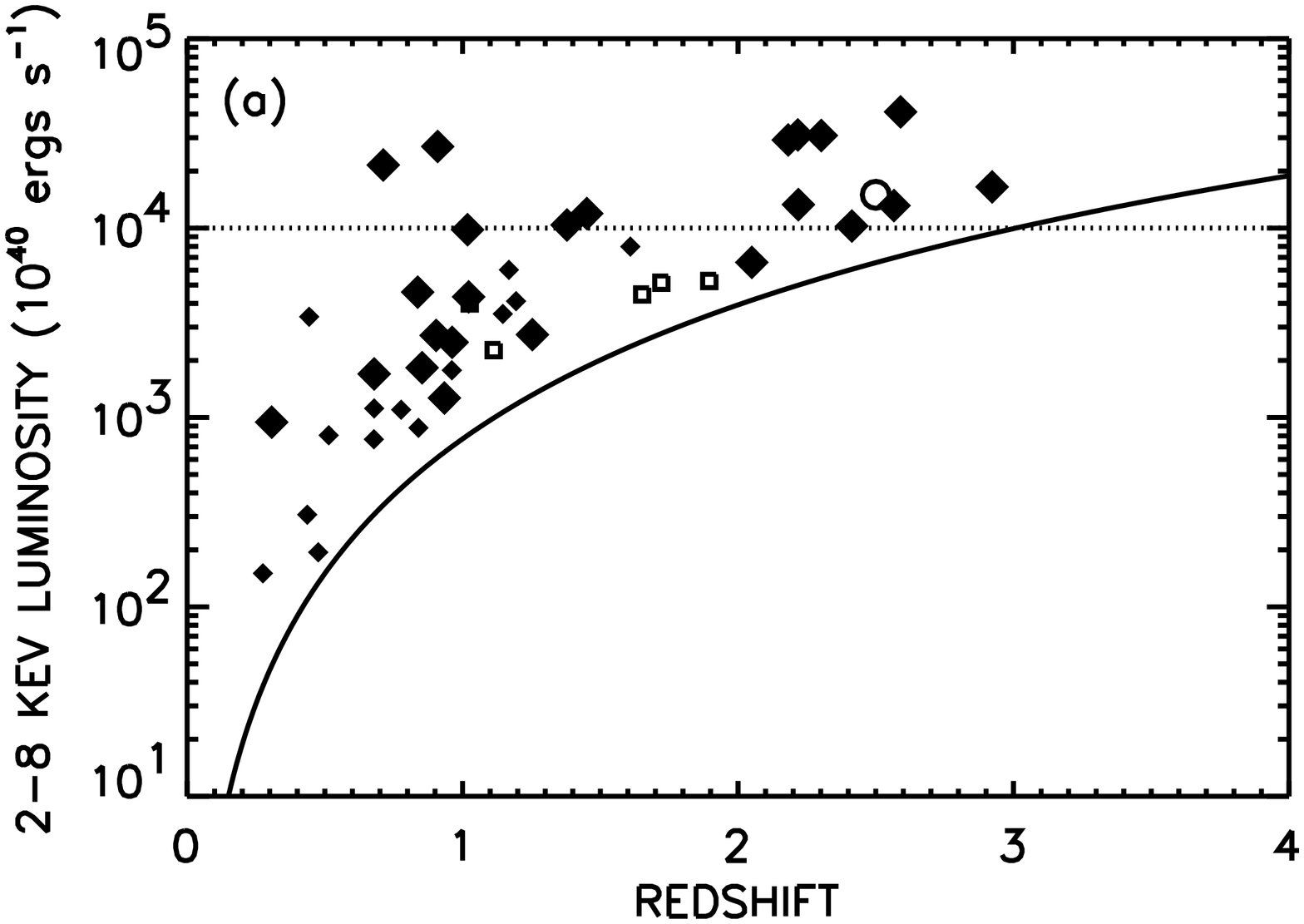,angle=90,width=3.5in}
\psfig{figure=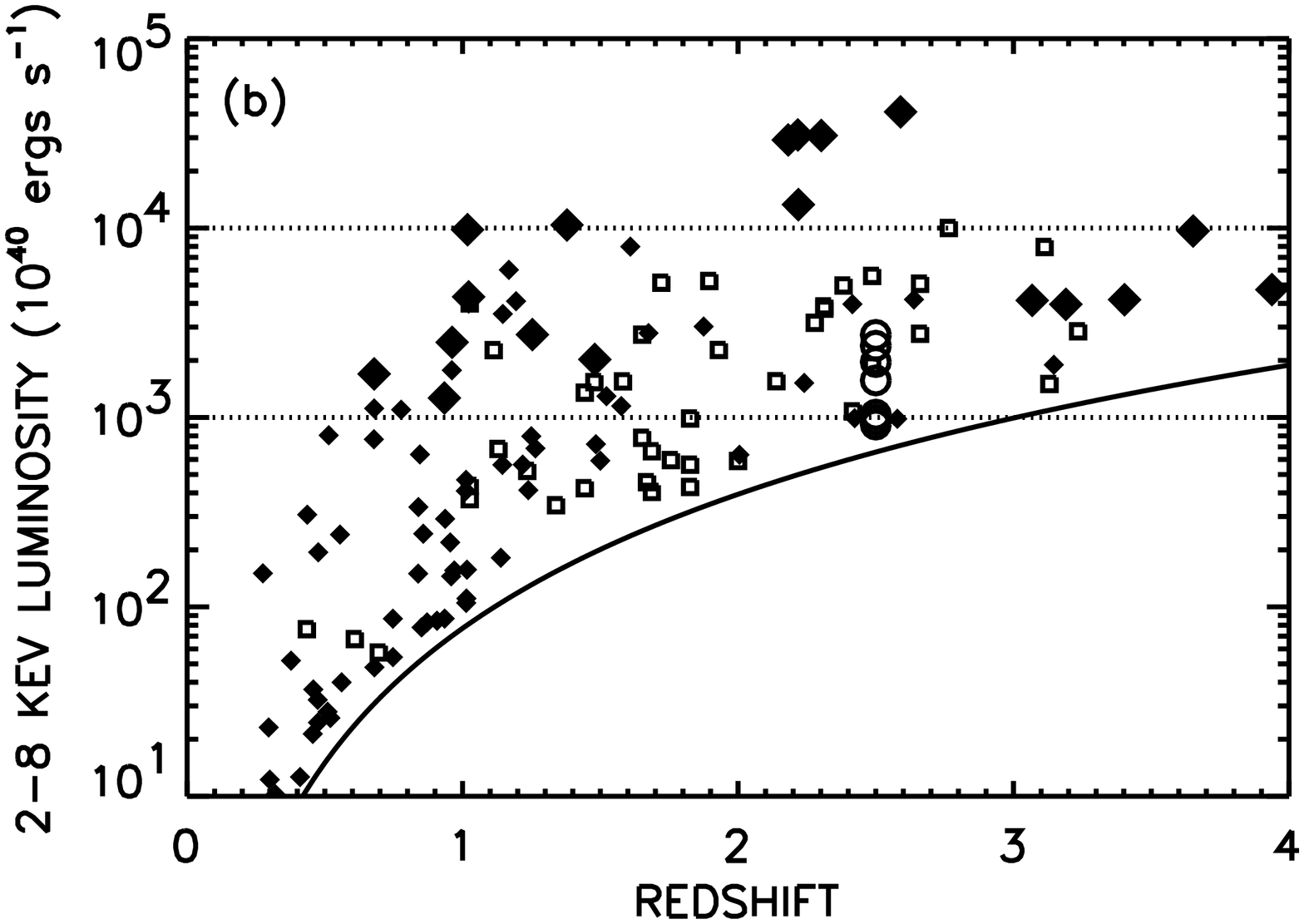,angle=90,width=3.5in}}
\vspace{6pt}
\figurenum{2}
\caption{
Rest-frame $2-8$~keV luminosity versus redshift for the CDF-N
X-ray sources in our (a) bright and (b) faint restricted subsamples.
The luminosities were determined using the observed-frame $0.5-2$~keV
X-ray fluxes and $K$-corrections calculated using a $\Gamma=1.8$
spectrum. Sources with spectroscopic redshifts are denoted by small
solid diamonds. Broad-line AGNs are denoted by large solid diamonds.
Sources with IR photometric redshifts are denoted by
open squares. The eight remaining sources without spectroscopic
or photometric redshifts (one in the bright subsample and seven
in the faint subsample) are denoted by open circles and
are shown at $z=2.5$. The solid curve in (a) shows the
bright subsample $0.5-2$~keV flux limit of
$1.7\times 10^{-15}$~ergs~cm$^{-2}$~s$^{-1}$ in a $10'$ radius region,
and the solid curve in (b) shows the faint subsample
$0.5-2$~keV flux limit of $1.7\times 10^{-16}$~ergs~cm$^{-2}$~s$^{-1}$
in an $8'$ radius region. The dotted horizontal line in (a) shows
the high-luminosity interval lower limit of $10^{44}$~ergs~s$^{-1}$,
and the dotted horizontal lines in (b) show the
intermediate-luminosity interval upper and lower limits
of $10^{44}$ and $10^{43}$~ergs~s$^{-1}$, respectively.
\label{fig2}
}
\end{figure*}

Having selected a highly complete X-ray sample, 
our next concern was that we sample the full luminosity ranges of 
interest without clipping out any sources. For the traditional 
intermediate and high X-ray luminosity
cut-offs of $L_{\rm 2-8~keV}=10^{43}$ and $10^{44}$~ergs~s$^{-1}$,
this very naturally sets the redshift interval to $z=2-3$.
We can see this from Figure~\ref{fig2}, where we show rest-frame
$2-8$~keV luminosity versus redshift for the CDF-N X-ray sources
in our (a) bright and (b) faint subsamples. 
We determined the rest-frame $2-8$~keV luminosities from the
observed-frame $0.5-2$~keV fluxes, assuming an intrinsic
$\Gamma=1.8$ spectrum. At $z=3$, observed-frame $0.5-2$~keV
corresponds to rest-frame $2-8$~keV, providing the best
possible match to lower redshift data. Moreover, the
$0.5-2$~keV {\em Chandra\/} images are deeper than the
$2-8$~keV images, so using the observed-frame soft X-ray
fluxes at high redshifts results in increased sensitivity.
We only show the luminosities at low redshifts for illustrative
purposes, since normally one would use the observed-frame
$2-8$~keV fluxes to calculate the $2-8$~keV luminosities at
these redshifts. The soft X-ray flux limits 
of our faint and bright subsamples {\em (solid curves)\/} 
correspond to rest-frame $2-8$~keV 
luminosities of $10^{43}$~ergs~s$^{-1}$ and $10^{44}$~ergs~s$^{-1}$, 
respectively, at $z=3$. Thus, we will not be clipping out any 
sources in either the $L_{\rm 2-8~keV}=10^{43}$ to 
$10^{44}$~ergs~s$^{-1}$ interval in the faint subsample, or 
the $L_{\rm 2-8~keV}=10^{44}$ to $10^{45}$~ergs~s$^{-1}$ 
interval in the bright subsample, if we choose $z=2-3$ for our 
analysis. 

%
%

Once we have chosen our soft X-ray flux and radius limits and 
our redshift interval, we can plot the spectroscopically identified 
$z=2-3$ sources in the flux-radius plane and see how many lie
in our restricted subsamples. This is shown in Figure~\ref{fig1}, 
where we denote with large squares the sources that have been 
spectroscopically identified to lie in the redshift interval 
$z=2-3$. The soft X-ray flux and radius limits for our two 
subsamples were chosen without regard to this distribution, as 
can be seen from the figure. There are 15 sources in our two 
subsamples with spectroscopic redshifts in the interval $z=2-3$.

\subsection{Probability of Misidentifications}
\label{secmisid}

Another issue we need to address is how many of the X-ray sources 
with detected optical/near-infrared (NIR) counterparts we may be 
misidentifying. Of the 160 sources in the 2~Ms exposure that lie 
within our two subsamples, 138 are detected above the $3\sigma$ 
limit of 23 in $H$. Thus, only 22 of the sources in our subsamples 
are not detected at the $3\sigma$ level in the NIR. 
By randomizing the source positions 
and remeasuring the $H$ band magnitudes many times, we found an 
11\% probability of misidentification. Thus, of the 138 sources 
in our subsamples with $3\sigma$ $H$ band detections, we would 
expect that about 15 might be contaminated by the projection of
random, superposed objects. 

Of the 22 sources that are not detected at the $3\sigma$ level
in $H$, 11 are detected in the IRAC $3.6\mu$m band at $<23$
magnitude, with a 4\% probability of misidentification based 
on randomized measurements. This corresponds to about one-half of
a source. Of the remaining 11 sources that are not detected at
$3.6\mu$m, 5 are detected at $R<25.5$, with a 10\% probability of 
misidentification based on randomized measurements. 
Again, this corresponds to about one-half of a source.

Thus, in total, we have optical/NIR
identifications for 154 of the 160 sources, eight of which
are stars, and we expect about 16 (or just over 10\%) of these 
to be false. Even if a large fraction of the 16 real sources 
suffering from random projections were in the $z=2-3$ range, 
which is not very likely, the correction to the number densities 
we derive subsequently would be small.

\section{Photometric Redshifts}
\label{secz}

Barger et al.\ (2002, 2003b)
computed photometric redshifts for the CDF-N X-ray sources using
broadband galaxy colors and the Bayesian code of Ben\'{\i}tez (2000).
They only used sources with probabilities for the photometric
redshift of greater than 90\%, resulting in about an 85\% success 
rate for photometric identifications. These redshifts are robust 
and surprisingly accurate (often to better than 8\% when compared 
with the spectroscopic redshifts) for non--broad-line AGNs.

%
%
\begin{figure*}[t]
\centerline{\psfig{figure=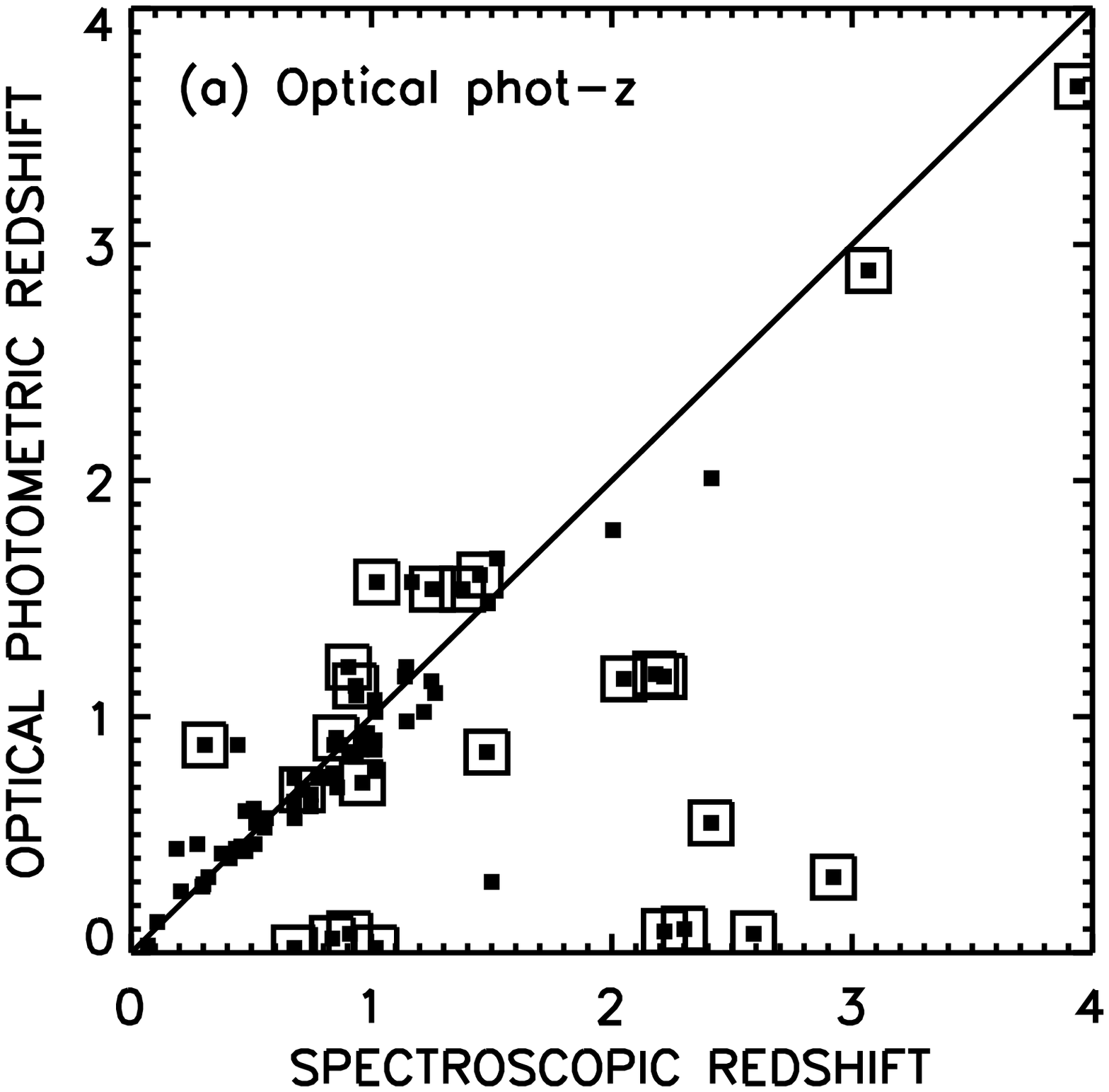,angle=90,width=3.5in}
\psfig{figure=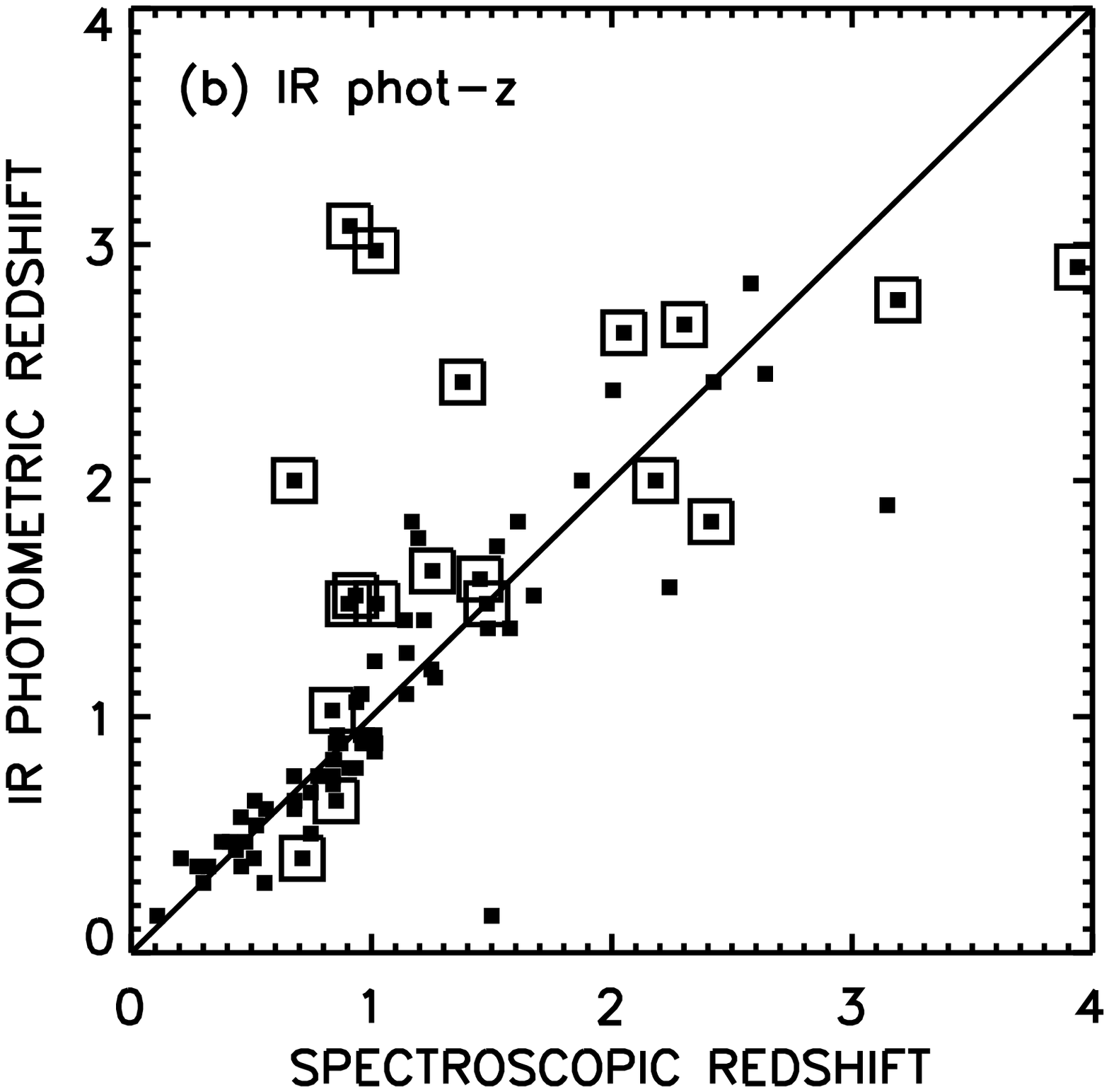,angle=90,width=3.5in}}
\vspace{6pt}
\figurenum{3}
\caption{
(a) Comparison of the optical photometric redshift estimates
with Bayesian probabilities $>0.90$ from Barger et al.\ (2003b)
with the spectroscopic redshifts.
(b) Comparison of the IR photometric redshift estimates from
this paper with the spectroscopic redshifts. Large open
squares denote broad-line AGNs. 
\label{fig3}
}
\end{figure*}

%
%
\begin{figure*}[b]
\centerline{\psfig{figure=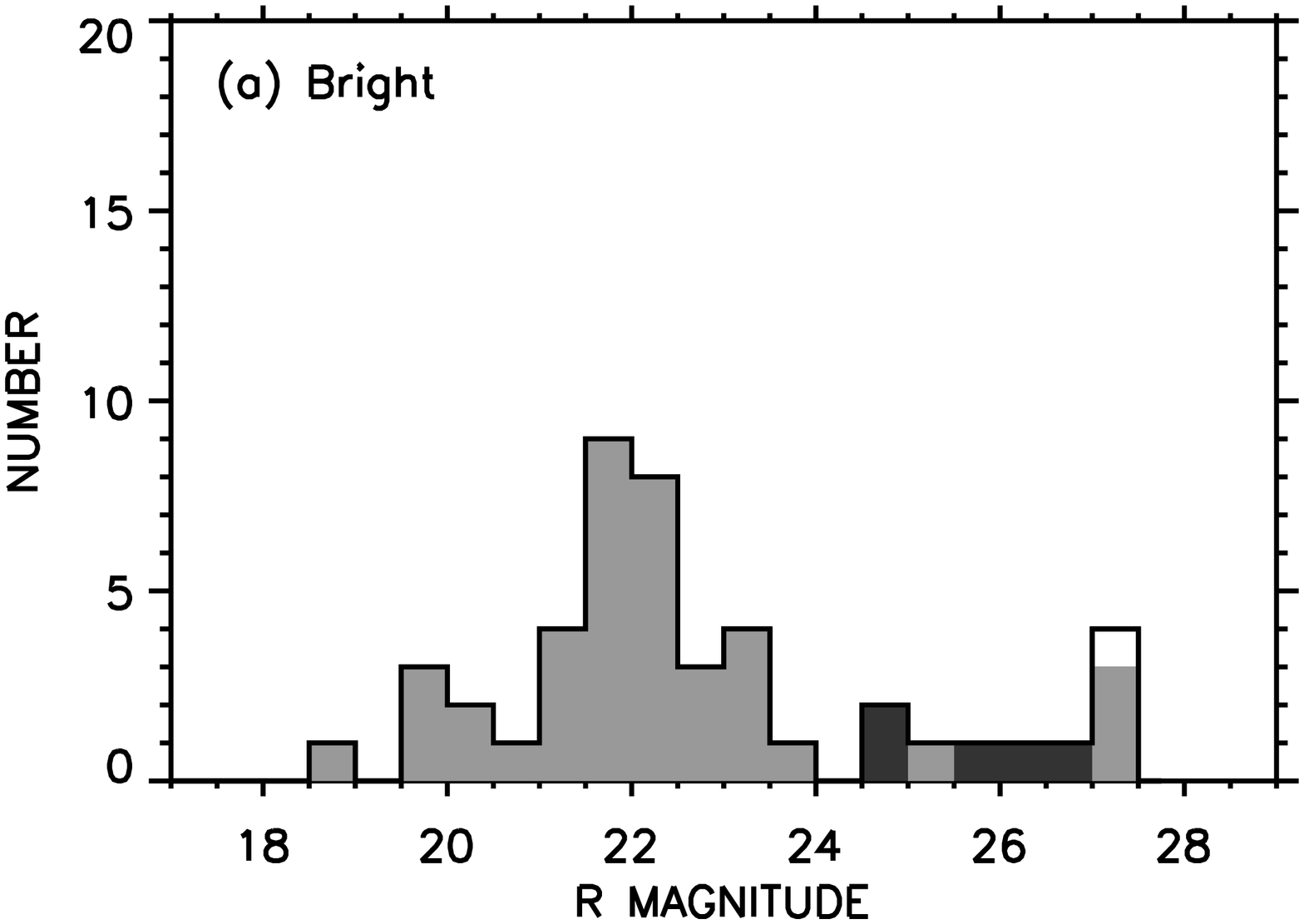,angle=90,width=3.5in}
\psfig{figure=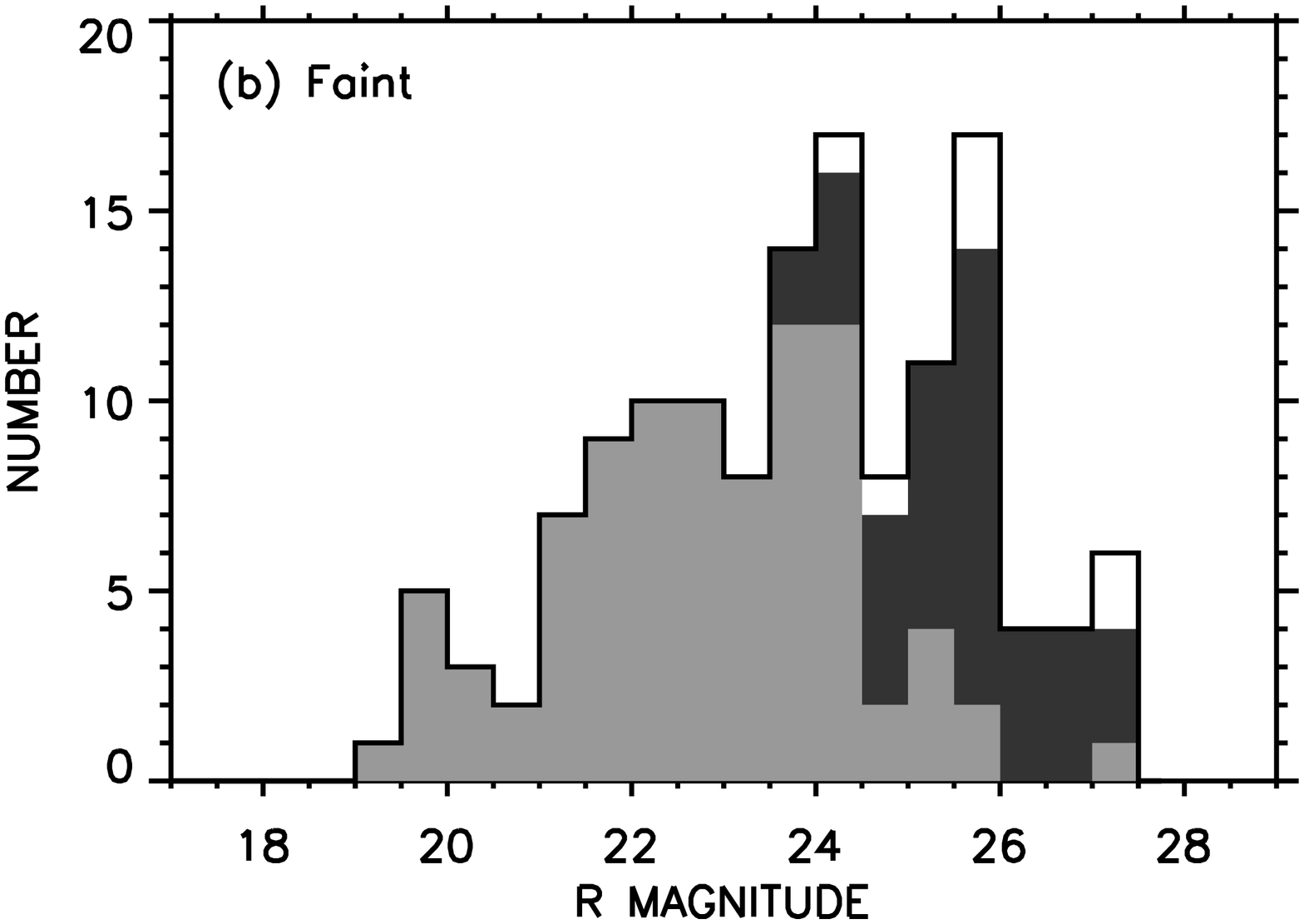,angle=90,width=3.5in}}
\vspace{6pt}
\figurenum{4}
\caption{
Fraction of the total {\em (open histogram)\/}
(a) bright and (b) faint restricted subsamples that
are spectroscopically {\em (light shading)\/} and
photometrically {\em (dark shading)\/} identified
vs. $R$ magnitude. All $R>27$ sources are placed
in the final magnitude bin.
\label{fig4}
}
\end{figure*}

%
%
\begin{figure*}[t]
\centerline{\psfig{figure=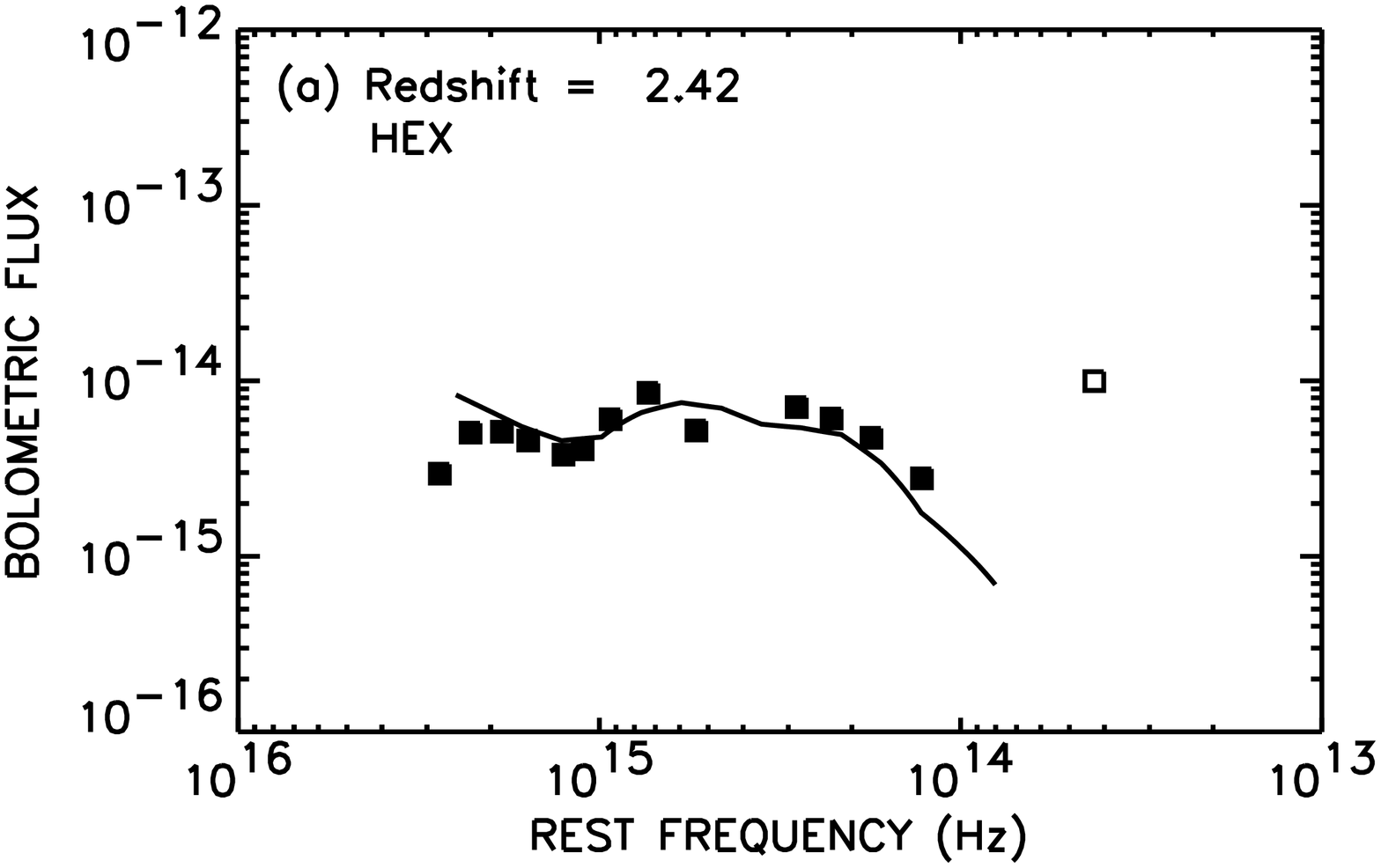,angle=90,width=3.5in}
\psfig{figure=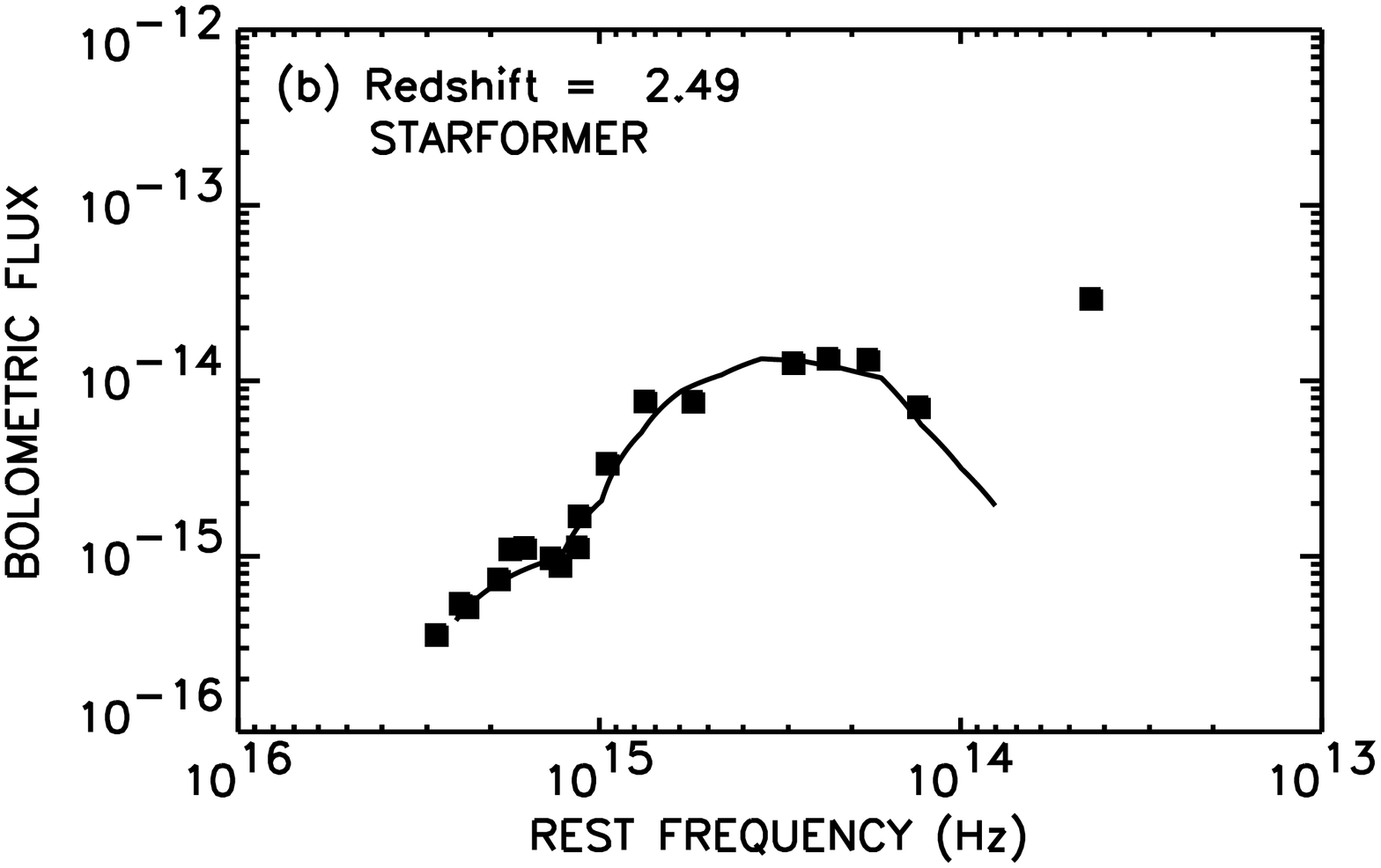,angle=90,width=3.5in}}
\centerline{\psfig{figure=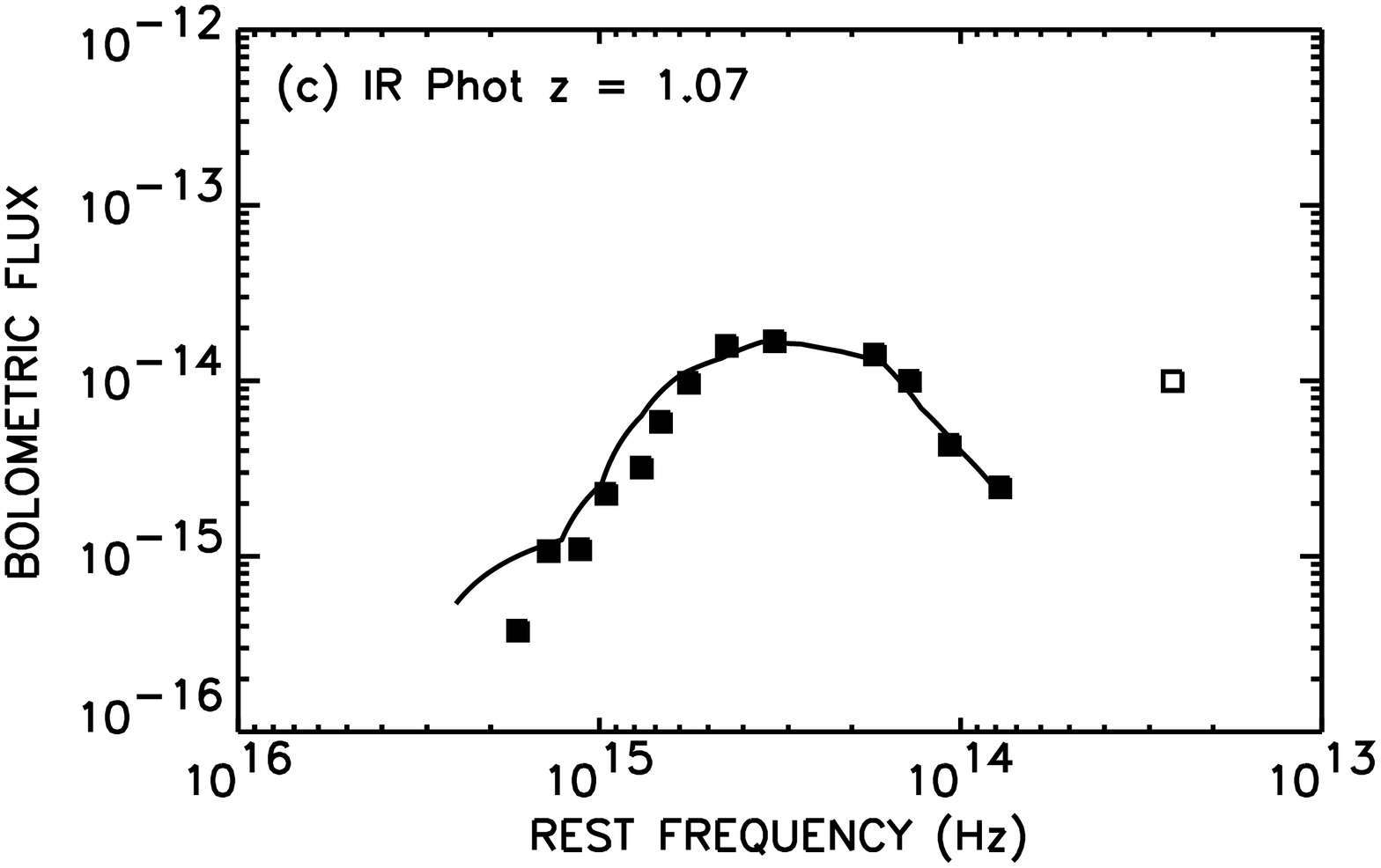,angle=90,width=3.5in}
\psfig{figure=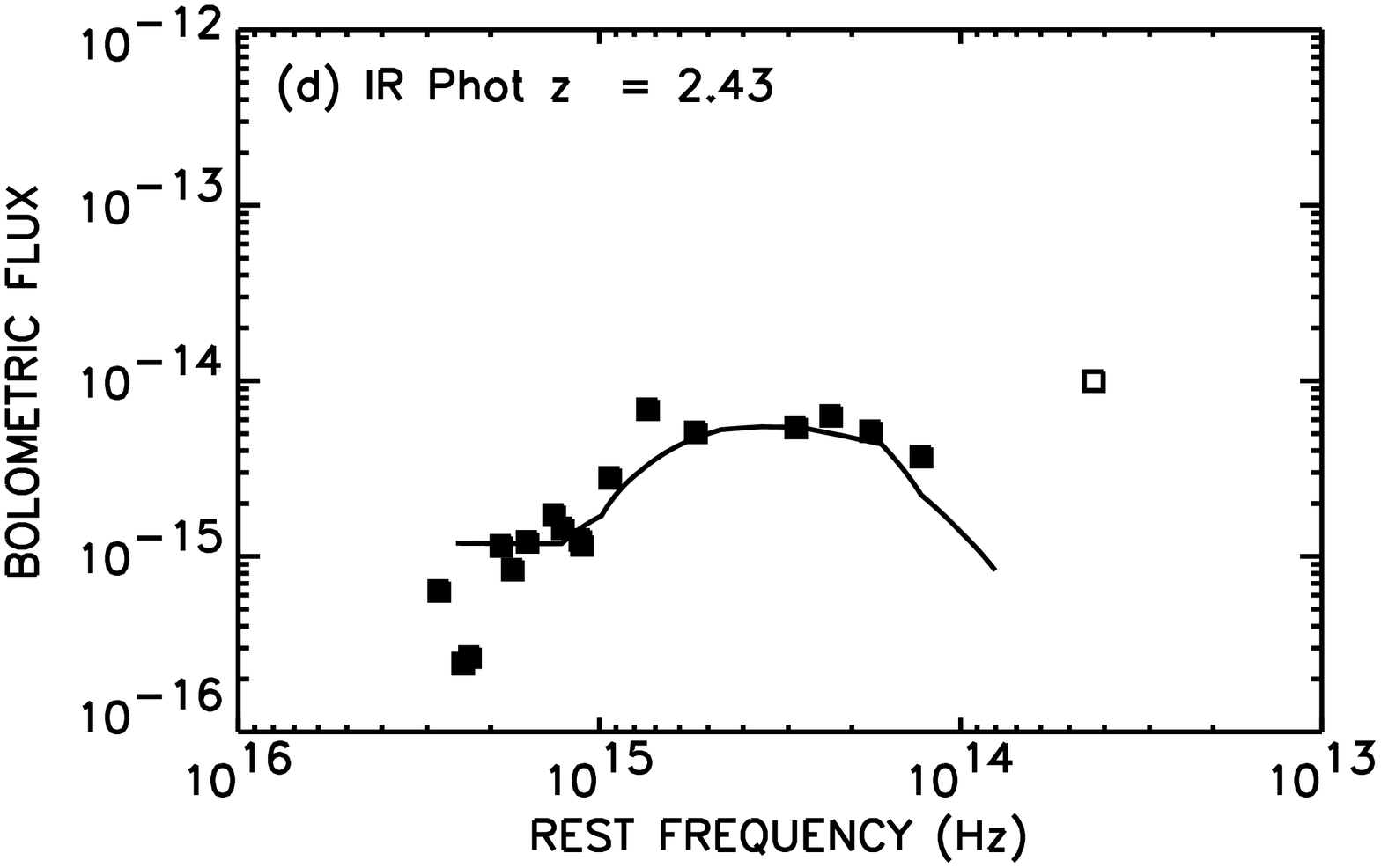,angle=90,width=3.5in}}
\vspace{6pt}
\figurenum{5}
\caption{
Bolometric flux ($\nu f_\nu$) vs. rest frequency for four
example X-ray sources in the CDF-N sample. Solid squares
show the optical ({\em HST\/} ACS and ground-based),
near-infrared ($J$ and $H$), and mid-infrared (3.6$\mu$m,
4.5$\mu$m, 5.8$\mu$m, 8.0$\mu$m, and 24$\mu$m)
measurements. Open squares denote $5\sigma$ upper limits.
Curves show the average spectral energy distributions of
the CDF-N X-ray sources of a given spectral class selected
as templates by the IR photometric redshift estimation
code (see text for details). Spectroscopic redshifts
and spectral classes are used in (a) and (b), and
IR photometric redshifts are used in (c) and (d).
\label{fig5}
}
\end{figure*}

We now have the advantage of the {\em Spitzer\/} 
IRAC $3.6\mu$m, $4.5\mu$m, $5.8\mu$m, and $8.0\mu$m data,
as well as the NIR $J$ and $H$ band data, for 
estimating photometric redshifts. 
Barger et al.\ (2005) classified the X-ray sources into 
four spectral classes: absorbers (i.e., no strong 
emission lines [EW([OII])$<3$~\AA\ or EW(H$\alpha+$NII)$<10$~\AA]), 
star formers (i.e., strong Balmer lines and no 
broad or high-ionization lines), high-excitation (HEX) sources 
(i.e., [NeV] or CIV lines or strong [OIII] 
[EW([OIII]~5007~\AA$)>3$~EW(H$\beta)$]), and broad-line AGNs
(i.e., optical lines having FWHM line widths
greater than 2000~km~s$^{-1}$). The measured spectral 
energy distributions (SEDs) of the X-ray sources in each of 
these spectral classes turn out to be remarkably tight. Thus, 
we can construct templates from the average SEDs for each class
and obtain an {\em IR photometric redshift\/} by fitting the 
templates to the optical through mid-infrared (MIR) data. 
We note that the template fitting is insensitive to the 
spectral class for sources without strong AGN signatures. 

In Figure~\ref{fig3}, we compare both (a) the optical photometric
redshifts from Barger et al.\ (2003), and (b) our new IR 
photometric redshifts with the spectroscopic redshifts.
Both photometric redshift techniques tend to fail on some
of the broad-line AGNs {\em (large open squares)\/}, but, fortunately, 
broad-line AGNs are straightforward to identify spectroscopically, 
even in the so-called redshift ``desert'' at $z\sim 1.5-2$, 
and nearly all of the 
CDF-N X-ray sources have now been spectroscopically observed. While 
the Bayesian optical photometric redshift technique gives slightly 
tighter values at low redshifts due to the presence of strong 
spectral features and breaks in the templates from 
Coleman, Wu, \& Weedman (1988) and Kinney et al.\ (1996), it does 
not do so well on the $z>2$ sources. In our subsequent analysis, 
we adopt the more robust IR photometric redshifts when a
spectroscopic redshift is not available. In Figure~\ref{fig4}, 
we show histograms of the spectroscopic {\em (light shading)\/} 
and IR photometric {\em (dark shading)\/} redshift 
identifications versus $R$ magnitude for the sources in our 
(a) bright and (b) faint subsamples.

The photometric redshifts not only tell us which of the
spectroscopically unidentified sources are likely to have
redshifts within the $z=2-3$ redshift interval, but they
also enable us to remove contaminants that are really at
lower redshifts. In Figure~\ref{fig5}, we show four example 
measured X-ray source SEDs {\em (solid squares)\/} redshifted
to the rest-frame using the printed spectroscopic or
IR photometric redshift. We have superimposed
on each SED the average spectral template chosen by the 
IR photometric redshift fitting for that source. We selected these 
four examples to illustrate (a, b) two spectroscopically identified 
sources with different spectral classes (also printed), both of which 
are in the redshift interval $z=2-3$, (c) one IR photometric 
redshift of a lower redshift source, and (d) one IR photometric 
redshift of a source in the redshift interval $z=2-3$.

Of the 44 sources with
$f_{0.5-2~{\rm keV}}>1.7\times 10^{-15}$~ergs~cm$^{-2}$~s$^{-1}$ 
within a $10'$
radius, 38 have spectroscopic redshifts. Of the remaining 6
sources, 5 have a robust IR photometric redshift, none
of which lie in the redshift interval $z=2-3$.
Of the 136 sources with 
$f_{0.5-2~{\rm keV}}>1.7\times 10^{-16}$~ergs~cm$^{-2}$~s$^{-1}$
within an $8'$ radius, 88 have spectroscopic redshifts.
Of the remaining 48, 41 have robust IR photometric redshifts, 
12 of which lie in the redshift interval $z=2-3$. That
leaves 8 sources in total with neither a spectroscopic
nor a photometric redshift; 6 were not clearly detected
in the $H$, IRAC $3.6\mu$m, or $R$ bands 
(see \S\ref{secmisid}), and the photometric redshifts are
not obvious from the SEDs for the remaining two.

\section{AGN number densities at $z=2-3$}
\label{secnum}

Six of the eight sources without spectroscopic or 
photometric redshifts could, if they were assigned redshifts 
at the center of the redshift interval $z=2-3$, have 
luminosities that would place them into one of our two luminosity 
intervals (one in the high-luminosity interval, and five in the 
intermediate-luminosity interval).
In Figure~\ref{fig2}, we denote the X-ray sources with spectroscopic 
redshifts by small, solid diamonds, and the broad-line AGNs by
large, solid diamonds. We use open squares to denote the sources 
with IR photometric redshifts, and we use open circles 
to denote the unidentified sources (these are plotted at $z=2.5$). 

%
%
\begin{figure*}[b]
\centerline{\psfig{figure=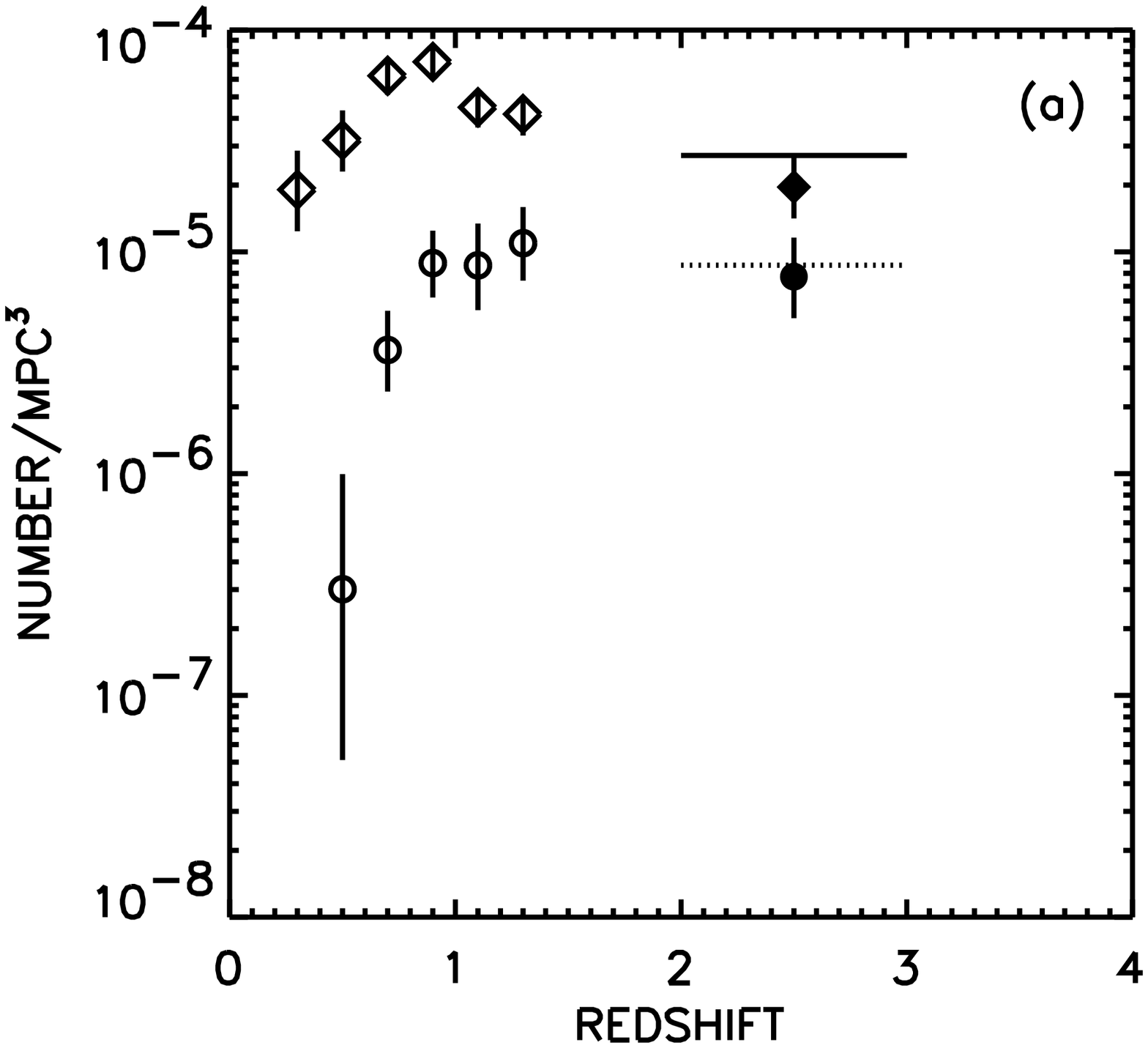,angle=90,width=3.5in}
\psfig{figure=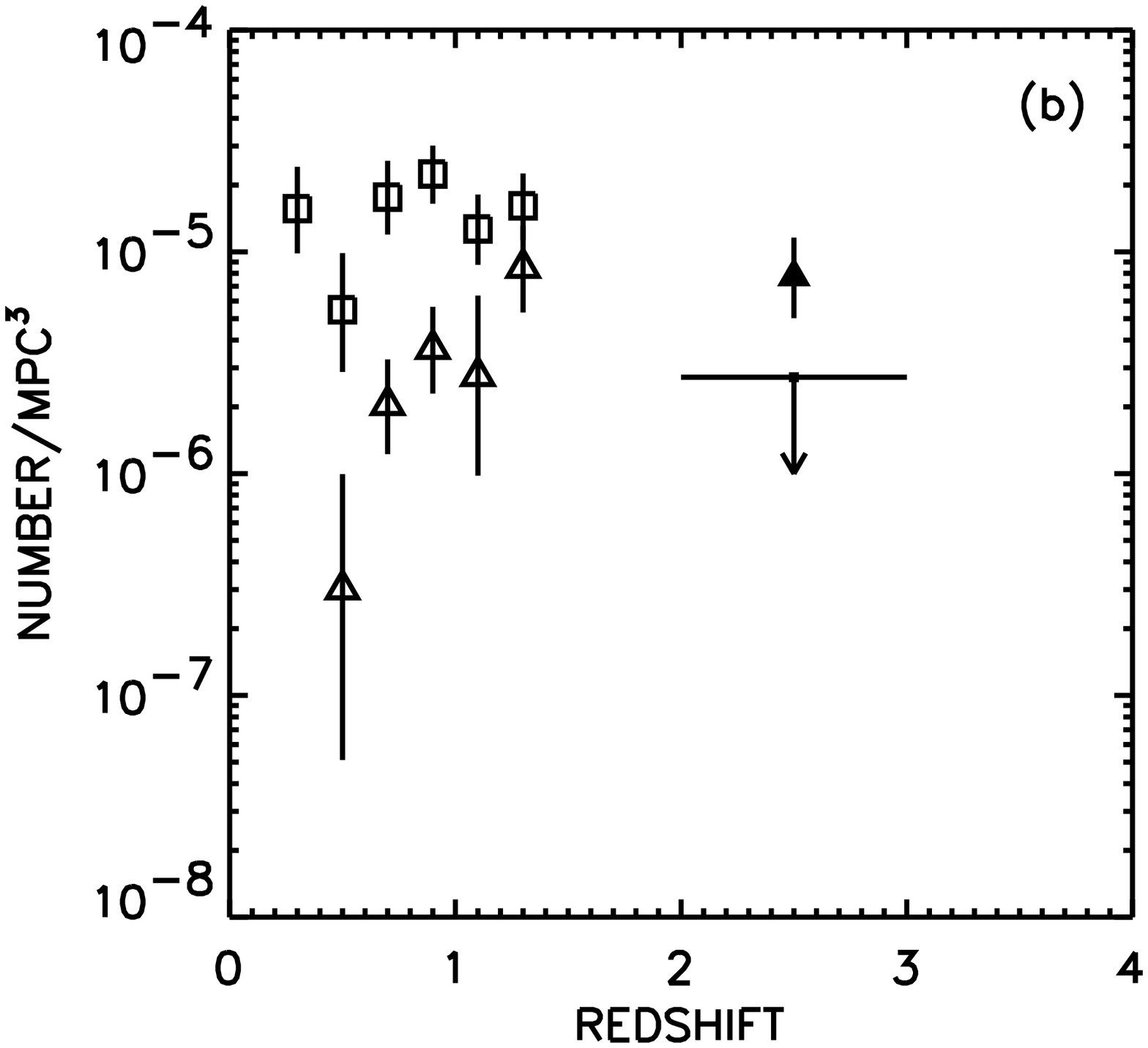,angle=90,width=3.5in}}
\vspace{6pt}
\figurenum{6}
\caption{
(a) Number density of sources with rest-frame $2-8$~keV
luminosities between $10^{43}$ and $10^{44}$~ergs~s$^{-1}$
{\em (diamonds)\/} and $10^{44}$ and
$10^{45}$~ergs~s$^{-1}$ {\em (circles)\/}.
The solid symbols were determined using sources with both
spectroscopic and IR photometric redshifts between $z=2-3$,
and the open diamonds and open circles at $z<1.5$ were
determined using the spectroscopic sample of Barger et al.\ (2005).
Points below (above) $z=2$ were determined from the
observed-frame $2-8$~keV ($0.5-2$~keV) sample. An
intrinsic $\Gamma=1.8$ was assumed, for which there is
only a small differential $K$-correction to correct
to rest-frame $2-8$~keV. Poissonian $1\sigma$ uncertainties
are based on the number of sources in each redshift interval.
The solid (dotted) horizontal bar shows the maximal number
densities in the $10^{43}-10^{44}$~ergs~s$^{-1}$
($10^{44}-10^{45}$~ergs~s$^{-1}$) interval found by assigning
redshifts of 2.5 to the 5 (1) unidentified sources that could
lie in these luminosity intervals at that redshift.
(b) Number density of broad-line AGNs with rest-frame $2-8$~keV
luminosities between $10^{43}$ and $10^{44}$~ergs~s$^{-1}$
{\em (squares)\/} and $10^{44}$ and $10^{45}$~ergs~s$^{-1}$
{\em (triangles)\/}. The symbols were all determined using
sources with spectroscopic broad-line AGN classifications
from Barger et al.\ (2005). The solid symbols were determined
for the $z=2-3$ interval. The solid square and horizontal bar
show the $1\sigma$ upper limit for the
$10^{43}-10^{44}$~ergs~s$^{-1}$ and $z=2-3$ interval, since
no broad-line AGNs were detected here.
\label{fig6}
}
\end{figure*}

We computed the AGN number densities at $z=2-3$ for the 
spectroscopically or photometrically identified sources 
in the rest-frame $2-8$~keV luminosity intervals 
$10^{43}-10^{44}$~ergs~s$^{-1}$
and $10^{44}-10^{45}$~ergs~s$^{-1}$ using the appropriate areas
and volumes. These number densities are shown in Figure~\ref{fig6}a 
as a solid diamond and a solid circle, respectively. 
The Poissonian $1\sigma$ uncertainties are based on the number
of sources in each redshift interval. Our high-luminosity sample 
consists of 8 sources in this redshift interval, all of which
are broad-line AGNs. Our intermediate-luminosity sample consists
of 13 sources in this redshift interval, none of which is a
broad-line AGN. 

We also computed upper limits on the number densities at $z=2-3$ 
by assigning redshifts of 2.5 to the six unidentified sources 
described above. We denote these upper limits by solid 
($10^{43}-10^{44}$~ergs~s$^{-1}$) and
dashed ($10^{44}-10^{45}$~ergs~s$^{-1}$) horizontal lines in 
Figure~\ref{fig6}a.

For comparison, we also plot in Figure~\ref{fig6}a the AGN
number densities at $z<1.5$ for our two luminosity intervals.
These number densities were determined from the spectroscopic sample
of Barger et al.\ (2005), which includes the CDF-N (Barger et al.\ 2003),
CDF-S (Szokoly et al.\ 2004), and CLASXS (Steffen et al.\ 2004) data.
We do not expect cosmic variance to be an issue on these
scalelengths (Yang et al.\ 2005).
We calculated the rest-frame $2-8$~keV luminosities for these
sources using the observed-frame $2-8$~keV band, assuming an
intrinsic $\Gamma=1.8$ spectrum. The Poissonian $1\sigma$ uncertainties
are again based on the number of sources in each redshift interval.
The Barger et al.\ (2005) spectroscopic sample is substantially 
complete at these redshifts and luminosities, and we expect that 
any incompleteness correction would be small.

In Figure~\ref{fig6}a, we see that the AGN number densities in 
both luminosity intervals show a steep rise at $z<1$.  The 
intermediate-luminosity number densities {\em (diamonds)\/}
then show a marked decline to $z=2-3$, while the high-luminosity 
number densities {\em (circles)\/}
remain relatively constant to $z=2-3$. If the number densities 
in the $L_{\rm 2-8~keV}=10^{43}-10^{44}$~ergs~s$^{-1}$ interval
were constant with redshift and equal to the peak value of 
$\sim 10^{-4}$~Mpc$^{-3}$ seen just below $z=1$ in this luminosity 
interval, then we would expect 66 sources in the $z=2-3$ 
interval at off-axis radii less than $8'$. In fact, we only have 
13 sources with spectroscopic and photometric redshifts in
this redshift interval in our faint subsample. Even if we were 
to add in all eight of the unidentified sources, this number would 
only rise to 21. Thus, we can reject the hypothesis that the 
number densities do not decline with redshift at above the 
$5\sigma$ level.

In Figure~\ref{fig6}b, we show the broad-line AGN number densities
for the same two luminosity intervals. The high-luminosity 
number densities show a dramatic rise at $z<1.5$
{\em (open triangles)\/}, and then a flattening to $z=2-3$
{\em (solid triangle)\/}, while the intermediate-luminosity 
number densities remain relatively constant at $z<1.5$
{\em (open squares)\/}, and then decline to $z=2-3$, where we 
only have an upper limit {\em (solid square and arrow)\/}. 
The $z<1.5$ behavior can be understood
in the context of the broad-line AGN luminosity functions given 
in Figure~18 of Barger et al.\ (2005). 
Over the $z=0-1.5$ redshift interval, the evolution of the 
broad-line AGN luminosity function is well defined by pure luminosity 
evolution with a rapid increase of luminosity with redshift. 
As a consequence, the $L_{\rm 2-8~keV}=10^{44}-10^{45}$~ergs~s$^{-1}$ 
interval, which lies on the steeply declining high-luminosity 
end of the luminosity function throughout this redshift interval, 
rises rapidly. In contrast, the 
$L_{\rm 2-8~keV}=10^{43}-10^{44}$~ergs~s$^{-1}$ interval
lies just above the peak luminosity in the luminosity
function at $z=0$, and the number densities in this
luminosity interval stay relatively constant
over the $z=0-1.5$ redshift interval.

\section{Comparison with Nandra et al.\ (2005)}
\label{secdisc}

N05 combined deep X-ray data of the CDF-N
(2~Ms) and of the Groth-Westphal Strip (GWS; 200~ks) with
the LBG surveys of Steidel et al.\ (2003) to estimate the
number density of intermediate-luminosity AGNs in the
interval $z=2.5-3.5$. Their method differs from the
X-ray follow-up surveys in that it uses a 
rest-frame UV-selected sample over a narrow range of redshift
with a cosmological volume determined from the optical selection
function. The X-ray data are only used to determine whether
an LBG hosts an AGN and to calculate an X-ray luminosity.
The Groth Strip data are much shallower than the CDF-N data, 
and they apply an empirically determined 20\% correction for 
X-ray incompleteness.

To compare our results directly with those of N05, we used our
CDF-N faint subsample to compute the AGN number densities at 
$z=2-3$ in the same rest-frame $2-8$~keV luminosity interval,
$L_{\rm 2-8~keV}=10^{43}-10^{44.5}$~ergs~s$^{-1}$, used by N05.
We note that by choosing to use a broader luminosity interval 
that goes to higher luminosities, N05 is pushing into the QSO 
regime, where we know the number densities continue to rise to 
higher redshifts than the $z=1$ redshift at which the
intermediate-luminosity number densities peak
(see Fig.~\ref{fig6}a). Our $z=2-3$ data point 
{\em (solid diamond in Fig.~\ref{fig7})\/} contains 17 sources, 
while the N05 data point contains 10.
We computed an upper limit on our $z=2-3$ number density by 
assigning redshifts of 2.5 to the five unidentified sources in
our faint subsample that could lie in this luminosity interval 
if they were at that redshift. We denote this upper limit by a 
solid horizontal line in Figure~\ref{fig7}.

N05 claimed that their number density estimate in the $z=1.5-3$
range was approximately an order of magnitude higher than that
of Cowie et al.\ (2003)---and roughly equal to the
Cowie et al.\ (2003) upper limit---and a factor of about 3
higher than that of Barger et al.\ (2005).
N05's comparison with Cowie et al.\ (2003) was based on the
$L_{\rm 2-8~keV}>10^{42}$~ergs~s$^{-1}$ evolution, and hence was
not shown on their Figure~1. They did show their estimates of the
Barger et al.\ (2005) space densities, which they obtained by
approximately fitting and then integrating the
Barger et al.\ (2005) luminosity functions over the  
$L_{\rm 2-8~keV}=10^{43}-10^{44.5}$~ergs~s$^{-1}$ range and then
accounting for the differences in the adopted cosmology in the
X-ray bandpass, but they did not show the upper limits from
Barger et al.\ (2005) on their figure.

For the purposes of a clear comparison between the different
analyses, we include on Figure~\ref{fig7} the Cowie et al.\ (2003)
data points {\em (solid squares)\/} recalculated for the
$L_{\rm 2-8~keV}=10^{43}-10^{44.5}$~ergs~s$^{-1}$ luminosity
interval and the N05 cosmology (which is the same cosmology used
throughout this paper).
We also show the upper limit given by Cowie et al.\ (2003) for
the $z=2-4$ interval {\em (dashed horizontal line)\/}.
We denote the N05 result by a solid circle.
We can see that there is no inconsistency between the
Cowie et al.\ (2003) points and the present work, nor between
the Cowie et al.\ (2003) upper limit and the N05 result.
However, our present work has tightened up the measured
number density and decreased the upper limit.

We also computed the $z<1.5$ AGN number densities in the
$L_{\rm 2-8~keV}=10^{43}-10^{44.5}$~ergs~s$^{-1}$ luminosity 
interval {\em (open diamonds in Figure~\ref{fig7})\/} using 
the spectroscopic sample of Barger et al.\ (2005). The decline 
in the AGN number densities at $z>1$ is again highly significant.
Since N05 reached a different conclusion, namely, that their
``$z=3$ estimate is consistent with, and supportive of,
the hypothesis that AGN activity remained roughly constant
between $z=1$ and $z=3$'', we have also included
on Figure~\ref{fig7} the Ueda et al.\ (2003) points
{\em (open triangles)\/} and limits {\em (dotted horizontal
lines)\/} that N05 compared with. The Ueda et al.\ (2003) values
are generally similar to the Barger et al.\ (2005) values, 
except at $z\sim 1$, where they have
a high point. This is a consequence of the fact that the 
Ueda et al.\ (2003) data at this redshift come almost entirely 
from the CDF-N data, which are known to contain a redshift sheet 
at a median redshift of $z=0.94$ (Barger et al.\ 2003). 
The Barger et al.\ (2005) number densities are smoothed out 
by the inclusion of the much wider CLASXS survey data 
(Steffen et al.\ 2004).

We measure a value of
$2.6^{+0.8}_{-0.6}\times 10^{-5}$~Mpc$^{-3}$ for the number density
of sources in the $L_{\rm 2-8~keV}=10^{43}-10^{44.5}$~ergs~s$^{-1}$
range in the $z=2-3$ interval, which may be compared with the
N05 value of $4.2^{+1.8}_{-1.4}$~Mpc$^{-3}$.
In both cases, the errors only reflect Poissonian uncertainties
from the small number of sources and are 68\% confidence limits
(Gehrels 1986). Although the N05 data point is statistically
consistent with the present number density, the larger errors
in N05 marginally permit a flat number density, while
the present data definitively rule it out.

%
%
\begin{inlinefigure}
\psfig{figure=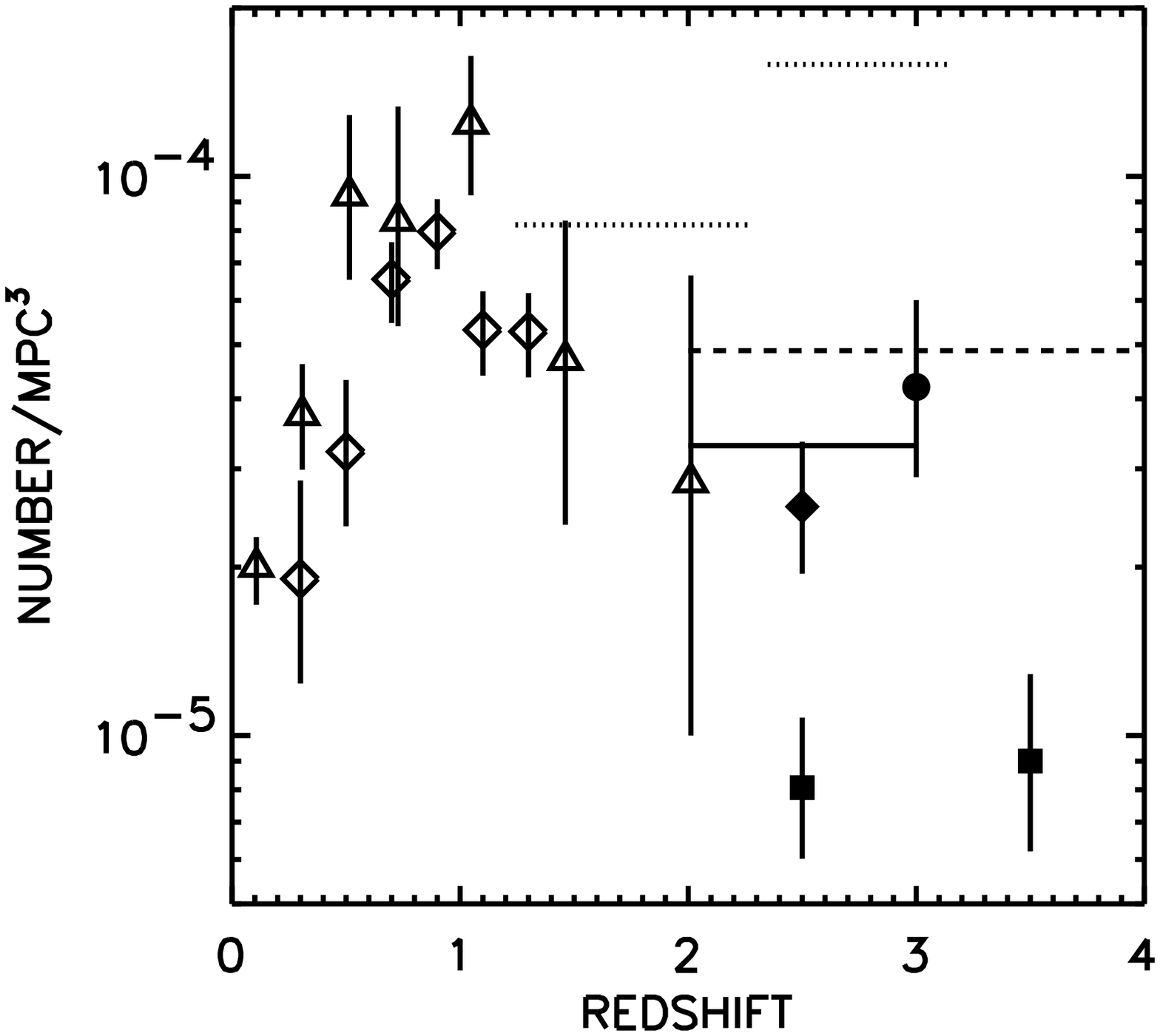,angle=90,width=3.5in}
\vspace{6pt}
\figurenum{7}
\caption{
Number density of sources with rest-frame $2-8$~keV
luminosities between $10^{43}$ and $10^{44.5}$~ergs~s$^{-1}$.
The solid diamond at $z=2-3$ was determined
using sources with both spectroscopic and IR photometric
redshifts, and the open diamonds at $z<1.5$ were determined
using the spectroscopic sample of Barger et al.\ (2005).
The solid horizontal bar shows the maximal number density
found by assigning redshifts of 2.5 to the 5 unidentified
sources that could lie in this luminosity interval at that
redshift. The solid circle at $z=3$ is from N05 from
their LBG surveys. The solid squares at $z=2.5$ and
$z=3.5$ are from the spectroscopic sample of Cowie et al.\ (2003).
The dashed horizontal line at $z=2-4$ shows the maximal
number density for the Cowie et al.\ (2003) sample, found by
assigning redshifts to all of the unidentified sources in their
sample that could lie in that redshift (and then luminosity) interval
redshifts at the center of the interval. The open triangles and
dotted horizontal bars show the Ueda et al.\ (2003) points
and upper limits, respectively. Poissonian $1\sigma$ uncertainties
are based on the number of sources in each redshift interval.
\label{fig7}
}
\addtolength{\baselineskip}{10pt}
\end{inlinefigure}

While the present data unambiguously show that 
the number densities at the intermediate luminosities
are declining to high redshifts, it should be noted that N05
were not correct in stating that a flat number density curve
would argue against previous suggestions that the majority of 
black hole accretion occurs at low redshifts, around $z=1$. 
The cosmological time available at high redshifts is much 
shorter than that at low redshifts, and the integral of the 
energy density production rate over the cosmic time from 
$z=0$ to larger $z$ would still be dominated by the sources 
around $z=1$ for the case of a number density which rises to 
$z=1$ and is constant at higher redshifts.

It is worth briefly discussing the three major advantages that
N05 argued their LBG method had over the X-ray follow-up method
for determining AGN number densities. First, they claimed that 
the optical LBG selection function is very well defined. However, 
as they noted in their paper, their volume element 
was calculated for galaxies, not AGNs, and the selection functions 
for both the broad-line AGNs and the narrow-line AGNs are likely 
to be quite different than that for the non-AGN LBGs, for which
they referenced Steidel et al.\ (2002) and Hunt et al.\ (2004).
They also noted that their method has the disadvantage 
of missing any AGNs that are too faint to be selected
in their optical survey and/or have colors that fail the
LBG selection criterion. Hunt et al.\ (2004) pointed out that
three of the X-ray sources with spectroscopic redshifts in the 
range $2.5<z<3.5$ reported by Barger et al.\ (2003b) were
not picked up in the Steidel et al.\ (2003) LBG survey, and
with the Barger et al.\ (2005) observations, that number has now 
doubled to six. N05 argued that this incompleteness should be 
accounted for in their effective volume calculation (they only 
expect to pick up about 40\% of the objects at $z=2.5-3.5$ 
compared to a top hat function), but given that this is such 
a large correction, the uncertainties on the optical LBG
selection function for AGNs are a concern.
Direct selection of X-ray sources avoids this problem and
should be much more robust.

Second, N05 argued that they can apply a lower X-ray detection 
threshold for their subsample of LBGs than can be applied when 
constructing a purely X-ray based catalog, thereby making their 
X-ray detections more complete. They declared this to be critical, 
since they are dealing with a population of sources which are 
close to the detection limit. However, since none of their 
additional five X-ray detected LBGs in the CDF-N (over and 
above the four LBGs detected in the CDF-N 2~Ms catalog by 
Alexander et al.\ 2003 and discussed in Nandra et al.\ 2002)
have luminosities $L_{\rm 2-8~keV}>10^{43}$~ergs~s$^{-1}$, this is 
not relevant to the issue of determining whether there is a
decline in the $L_{\rm 2-8~keV}=10^{43}-10^{44.5}$~ergs~s$^{-1}$ range.

Finally, N05 stated that their LBG color selection mostly avoids 
any concerns about spectroscopic incompleteness, since the 
probability that the non-spectroscopically identified LBGs are 
$z\sim 3$ galaxies is about 96\%. The high spectroscopic and 
photometric completeness of our $z=2-3$ X-ray sample mitigates 
this issue for our current analysis. 

Probably the most important concern about the N05 methodology 
is that they must correct for two window functions (a very 
incomplete optical LBG selection function that is not well 
understood for AGNs, and an X-ray selection function) rather
than one. In order to determine whether their 
optical LBG selection function is the same for galaxies with 
substantial AGN contributions as it is for those without,
they would need to spectroscopically identify a complete X-ray 
sample. However, once they had undertaken such a spectroscopic 
survey to calibrate their optical LBG selection, then there 
would be no need for them to redo the AGN number densities 
using the LBG method.

\section{Summary}
\label{secsum}

In summary, we constructed two uniform, X-ray flux-limited, 
highly spectroscopically complete subsamples of 160 sources
(eight of which are stars) with negligible X-ray incompleteness 
from the CDF-N 2~Ms X-ray data to compute AGN 
number densities in the $z=2-3$ redshift interval for both high 
($L_{\rm 2-8~keV}=10^{44}-10^{45}$) and intermediate 
($L_{\rm 2-8~keV}=10^{43}-10^{44}$) rest-frame $2-8$~keV
luminosity intervals. Of the 152 non-stellar sources, 102
are spectroscopically identified. In this paper, we used new 
UH~2.2m ULBCAM $J$ and $H$ band imaging and {\em Spitzer\/} 
IRAC imaging to estimate IR photometric redshifts for the 
sources without spectroscopic redshifts, increasing our 
identified non-stellar sample to 144. Our final galaxy 
subsamples contain no more than eight unidentified sources. 

We then calculated the $z=2-3$ AGN number densities (for all spectral
types together and for broad-line AGNs alone) for both luminosity
intervals and compared them with those at $z<1.5$, which we 
calculated from the spectroscopic sample of Barger et al.\ (2005). 
We find clear evidence for a decrease in the 
$L_{\rm 2-8~keV}=10^{43}-10^{44}$ AGN number densities at $z>1$
and can reject the hypothesis that the number densities remain
flat to $z=2-3$ at above the $5\sigma$ level.

\acknowledgements
We thank the referee and Paul Nandra for helpful comments that
improved the manuscript. We gratefully acknowledge support from 
NSF grants AST 02-39425 (A.J.B.) and AST 04-07374 (L.L.C.),
the University of Wisconsin Research Committee 
with funds granted by the Wisconsin Alumni Research Foundation, 
the Alfred P. Sloan Foundation, and the David and Lucile Packard
Foundation (A.J.B.).

\end{document}